\input{aipcheck}

\documentclass[final]{aipproc}

\layoutstyle{8x11double}

\def\gta{\ifmmode {\mathbin{\lower 3pt\hbox   %> or of order
    {$\,\rlap{\raise 5pt\hbox{$\char'076$}}\mathchar"7218\,$}}}
    \else {${\mathbin{\lower 3pt\hbox
    {$\rlap{\raise 5pt\hbox{$\char'076$}}\mathchar"7218\,$}}}
    $}\fi}
\def\lta{\ifmmode {\,\mathbin{\lower 3pt\hbox   %< or of order
    {$\,\rlap{\raise 5pt\hbox{$\char'074$}}\mathchar"7218\,$}}}
    \else {${\mathbin{\lower 3pt\hbox
    {$\rlap{\raise 5pt\hbox{$\char'074$}}\mathchar"7218\,$}}}
    $}\fi}

\begin{document}

\title{Interpreting QPOs from Accreting Neutron Stars}

\author{M. Coleman Miller}{
  address={University of Maryland at College Park}
}

\begin{abstract}
The high time resolution and large area of the Rossi X-ray Timing
Explorer have been essential in the detection and characterization of
high-frequency quasi-periodic variability in the flux from neutron
stars in low-mass X-ray binaries.  An unknown phenomenon prior to
RXTE, kilohertz quasi-periodic oscillations (QPOs) have now been
detected from more than twenty systems.  Their high frequencies (up
to 1330 Hz) imply that they are generated close to the neutron star,
where general relativistic effects are expected to play an important
role.  I summarize current models for the kilohertz QPO
phenomenon.  In particular, I show that there is a significant
domain of agreement among the models that can be used to constrain
neutron star structure and look for signatures of highly curved
spacetime in the properties of the QPOs.

\end{abstract}

\maketitle

%%%%%%%%%%%%%%%%%%%%%%%%%%%%%%%%%%%%%%%%%%%%
%% MAINMATTER
%%%%%%%%%%%%%%%%%%%%%%%%%%%%%%%%%%%%%%%%%%%%

\section{Introduction}

Neutron stars are important laboratories for physics at
high densities.  Unlike the matter in relativistic heavy-ion colliders,
the matter in the cores of neutron stars has a thermal energy that
is much less than its rest-mass energy.  Various researchers have
speculated whether neutron star cores contain primarily nucleons,
or whether degrees of freedom such as hyperons, quark matter, or
strange matter are prevalent (see Lattimer \& Prakash 2001 for a
recent review of high-density equations of state).  In addition,
the strongly curved spacetime around neutron stars implies
that we could observe predicted effects of strong gravity,
such as frame-dragging or signatures of an innermost stable
circular orbit.  

The fast timing phenomena observed from accreting neutron stars
with the {\it Rossi} X-ray Timing Explorer (RXTE) offer outstanding
opportunities for us to probe the regimes of high density and strong
gravity.  In particular, the kilohertz quasi-periodic brightness
oscillations (kHz QPOs) observed from more than twenty neutron stars
in low-mass X-ray binaries are promising, because their high frequencies
imply an origin near the star, where general relativity must play a
role.  

Here we discuss briefly some of the proposals for the origin of these
QPOs, and their implications for strong gravity and dense matter.
Although currently no first-principles magnetohydrodynamic simulations
produce sharp QPOs (a situation expected to change in the next few
years as computers become faster and more effects can be included), 
we show that current observational constraints on
models are significant enough to allow fairly confident inferences.
In \S~2 we show that general relativity must inevitably influence
the QPO phenomenon, independent of any detailed models.  In \S~3
we describe the constraints on models that follow from the observations,
and the constraints on stellar mass and radius that follow from the
viable options.  We conclude in \S~4 by discussing what discoveries
about dense matter and strong gravity can be expected with a future
$\sim 10$~m$^2$ timing instrument.

\section{The Effects of General Relativity}

Some of the most exciting potential implications of QPOs involve
the effects of general relativity.  These include signatures of
unstable circular orbits, and general relativistic frame-dragging.
However, if general relativity does not influence QPOs then the
interpretations and implications are less clear.

From the observational standpoint, a potential link to nonrelativistic
systems was made by Mauche (2002), following the work of Psaltis,
Belloni, \& van der Klis (1999).  These authors have shown that
particular pairs of QPOs from sources (sometimes selected from more than
two QPOs in a given system) follow a trend that, on a log-log plot,
appears to link neutron star systems with black hole and white dwarf
systems.  Since white dwarfs are not significantly relativistic, Mauche
(2002) concludes that the phenomenon as a whole  cannot involve general
relativity, and hence favors a model such as the one proposed by
Titarchuk and collaborators, in which the QPOs arise from classical disk
oscillations (e.g., Titarchuk \& Osherovich 1999; Osherovich \&
Titarchuk 1999; Titarchuk \& Osherovich 2000; Titarchuk 2002, 2003).
Abramowicz and Klu\'zniak have also proposed a mechanism of nonlinear disk
resonances to explain black hole systems (Abramowicz \& Klu\'zniak 2001;
Abramowicz et al. 2003a), which they suggest could extend to neutron
stars as well (Abramowicz et al. 2003b).

There are still many unknowns about the brightness oscillations from  disks
in binary systems, and it could be that there are some underlying
mechanisms in common between black holes, neutron stars, and white dwarfs.
However, whatever the detailed mechanism is, basic physical considerations
require that general relativity will have an impact near neutron stars or
black holes.  For example, consider the highest frequency brightness
oscillation ever observed, $\nu_{\rm QPO}=1330$~Hz, from 4U~0614+091 (van
Straaten et al. 2000).  The frequency of the innermost stable circular
orbit is $\lta 1500$~Hz for $M\gta 1.6\,M_\odot$ (see below).  This means
that all frequencies (e.g., orbital, vertical epicyclic, and radial
epicyclic) are altered by the curved spacetime.  In turn,  oscillation
modes of the disk are altered as well.  Even radiative transfer is affected
subtlely, by light deflection effects.  Thus, regardless of the underlying
mechanism, general relativity will alter the basic picture.

One can also look to observations to find that there are important
differences between the types of sources.  Abramowicz \& Klu\'zniak
(2001) made the prescient suggestion that for black hole QPOs there
would be small integer ratios between frequencies, before the first
detections of such a ratio (see Table~\ref{tab:ratio}).  There may also
be sources with small integer frequency ratios different than 3:2
(e.g., the possible 4:1 ratio found in 4U~1630--47 by Klein-Wolt,
Homan, \& van der Klis 2003).  This insight
may well prove to be a key to understanding this phenomenon.  However,
neutron star systems do not have a similar preference for any particular
frequency ratio.  For example, the most recent data for Sco X-1 (kindly
provided by Mariano M\'endez) are plotted in Figure~1.  As can be seen,
the 3:2 ratio that is prominent in black hole sources is not evident
here (the apparent clustering found by Abramowicz et al. 2003b used
older, less precise data for Sco X-1).

\begin{table}
\begin{tabular}{lrr}
\hline
   \tablehead{1}{l}{b}{Source}
  & \tablehead{1}{r}{b}{$\nu_{\rm upper}$ (Hz)}
  & \tablehead{1}{r}{b}{$\nu_{\rm upper}/\nu_{\rm lower}$}\\
\hline
GRS~1915+105\tablenote{Strohmayer 2001b} & 67$\pm$5 
& 1.63$\pm$0.13\\
XTE~J1550--564\tablenote{Miller et al. 2001; Remillard et al. 2002} & 
272$\pm$20 & 1.48$\pm$0.24\\
GRO~J1655--40\tablenote{Strohmayer 2001a; Remillard et al. 2002} & 
450$\pm$20 & 1.5$\pm$0.13\\
\hline
\end{tabular}
\caption{Upper peak frequencies and frequency ratios for black hole
QPO pairs.  GRS~1915+105 may also have a pair at $\approx$168~Hz and
$\approx$113~Hz (e.g., Remillard \& McClintock 2003).}
\label{tab:ratio}
\end{table}

\begin{figure}
  \includegraphics[height=.3\textheight]{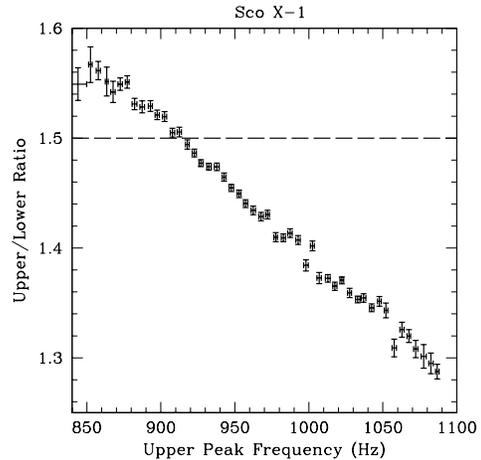}
  \caption{Ratios of upper to lower QPO peak frequencies for the
neutron star LMXB Sco~X-1 (data kindly provided by Mariano M\'endez).
The location of a 3:2 ratio is indicated by the dashed line.  Unlike
in the case of black hole QPO pairs, there is no preference in neutron
star sources for a particular ratio or set of ratios.  Similar trends
are evident in other neutron star sources.}
\end{figure}

In addition, there is strong evidence that the spin frequency plays a role
in generating at least one of the two strong kHz QPOs observed in neutron
stars.  This evidence comes from a comparison of the frequency separation
$\Delta\nu$ between kHz QPOs with the  spin frequency $\nu_{\rm spin}$, in
cases where both are known.  In  Table~\ref{tab:spin} we compare the range
of observed frequency separations with the spin frequency as inferred from
the persistent oscillations in SAX~J1808--3658 (Wijnands \& van der Klis
1998; Chakrabarty \& Morgan 1998) and from the frequency of burst
brightness oscillations in the other sources.  The identification of the
frequency of burst brightness oscillations with the spin frequency (or
close to  it) is well-established by observations of burst oscillations in
SAX~J1808--3658 (Chakrabarty et al. 2003) and XTE~J1814--338 (Strohmayer
et al. 2003) that are extremely close to the frequency of the oscillations
during persistent emission.  From  Table~\ref{tab:spin} we see that all
sources have $\Delta\nu$ close to $\nu_{\rm spin}$ or $\nu_{\rm spin}/2$,
and indeed 4U~1702--429, SAX~J1808--3658, and KS~1731--260 have
$\Delta\nu$ consistent with exactly $\nu_{\rm spin}$ or $\nu_{\rm
spin}/2$, within the uncertainties.  As indicated by Table~\ref{tab:spin},
the frequency difference does vary in several sources, requiring
modification of basic spin modulation ideas (see, e.g., Lamb \& Miller
2001), but the importance of the spin is clear. 

The involvement of the spin, e.g., through radiation effects or stellar
magnetic fields, indicates that there are at least some processes
affecting neutron star QPOs that are different from those that generate
black hole QPOs.  No such relation is evident for white dwarfs (Mauche
2002), suggesting again that even if there is a master underlying
mechanism at work, there are important differences between the different
classes of objects. General relativity does play a role for neutron stars
and black holes, hence full understanding of the QPOs is promising for our
understanding of strong gravity effects.

\begin{table}
\begin{tabular}{lrr}
\hline
   \tablehead{1}{l}{b}{Source}
  & \tablehead{1}{r}{b}{$\nu_{\rm spin}$ (Hz)}
  & \tablehead{1}{r}{b}{$\Delta\nu$ (Hz)}\\
\hline
4U~1916--053\tablenote{Boirin et al. 2000; Galloway et al. 2001} & 270 
& 290--348\\
4U~1702--429\tablenote{Markwardt, Strohmayer \& Swank 1999} & 329 
& 333$\pm$5\\
4U~1728--34\tablenote{Strohmayer et al. 1996} & 363 & 342--363\\
SAX~J1808--3658\tablenote{Wijnands et al. 2003; Chakrabarty et al.
2003} & 401 & 195$\pm$6\\
KS~1731--260\tablenote{Smith, Morgan, \& Bradt 1997; Wijnands \&
van der Klis 1997} & 524 & 260$\pm$10\\
4U~1636--536\tablenote{Di Salvo, M\'endez, \& van der Klis 2003; Jonker,
M\'endez, \& van der Klis 2002; M\'endez, van der Klis, \& van Paradijs 1998;
Wijnands et al. 1997} & 581 & 250--323\\
4U~1608--52\tablenote{Berger et al. 1996; M\'endez et al. 1998; M\'endez
et al. 1999; Yu et al. 1997} & 620 & 225--313\\
\hline
\end{tabular}
\caption{Spin frequency and frequency separation for neutron star
LMXBs.  4U~1702, SAX~J1808, and KS~1731 have single measurements of 
$\Delta\nu$, with uncertainties indicated.}
\label{tab:spin}
\end{table}

As an aside, the demonstration that $\Delta\nu\approx \nu_{\rm spin}/2$
in SAX~J1808--3658 is an indication
of how the superb data available from RXTE, combined with sophisticated
analysis (e.g., Chakrabarty et al. 2003, Wijnands et al. 2003), are
still facilitating qualitative leaps in our understanding.  Prior to the
SAX~J1808 analysis, I and many other researchers argued that 
$\Delta\nu\approx\nu_{\rm spin}$ in all cases.  The new results have 
forced modifications of the original
models (for a recent proposal, see the sonic point and spin resonance
beat frequency model of Lamb \& Miller 2003), proving again the
importance of high-quality timing data.

\section{Implications of QPOs}

As discussed in, e.g., Lamb \& Miller (2003), observations of kHz QPOs
in neutron star LMXBs give a number of clues to their physical origin.

\begin{itemize}

\item The spin frequency is involved in producing the observed frequency
      differences.  The difference can be close to $\nu_{\rm spin}$ or
      $\nu_{\rm spin}/2$.  So far, sources with $\nu_{\rm spin}>400$~Hz
      always have $\Delta\nu\approx\nu_{\rm spin}/2$, whereas sources with
      $\nu_{\rm spin}<400$~Hz always have $\Delta\nu\approx\nu_{\rm spin}$
      (Muno et al. 2001).

\item This appears to be a single sideband phenomenon.  That is, if the
      spin frequency modulates some other frequency, only one additional
      strong QPO is produced (additional QPOs have been found in 4U~1608--52, 
      4U~1728--34, and 4U~1636--536 by Jonker, M\'endez, \& van der Klis
      2000, but these are much weaker than the primary peaks).  This
      restricts models significantly; for example, amplitude modulation
      of one frequency by another produces two sidebands of equal strength.

\item An excellent candidate for the other frequency is the orbital frequency
      or something close to it.  The requirements are that the frequency be
      in the right range, while also being able to
      change frequency by hundreds of Hertz (for a review of the observational
      properties see, e.g., van der Klis 2000 or J. Swank, these proceedings).
      The orbital frequency has these properties.  If the orbital frequency
      is involved, then because one expects accretion to align the stellar
      spin with the sense of the accretion disk over a time short compared
      to the accretion lifetime, the orbital and spin directions are
      the same and hence the orbital frequency is expected to be close to
      the {\it upper} peak frequency.

\end{itemize}

Detailed models need to identify a mechanism that produces the
QPOs, selects a particular orbital radius
among many, and allows this radius to change significantly (as indicated by
the changing QPO frequencies).  The current leading candidates include some
variant of a beat frequency model (e.g., Lamb \& Miller 2003), or possibly
a resonance with the spin, modulating other frequencies (D. Psaltis,
presented at the ``Neutron Stars on Fire" conference, Princeton, NJ, 11-13
May 2003).  In the former case, the upper peak frequency $\nu_{\rm upper}$
is identified with a frequency close to an orbital frequency $\nu_{\rm
orb}$ at some special radius (e.g., the sonic radius; see
Miller, Lamb, \& Psaltis 1998), and in
the latter case it is identified with a vertical epicyclic frequency of a
nearly circular orbit $\nu_{\rm vertical}$ (e.g., Abramowicz et al.
2003a).  For constraints on neutron star structure these amount to the same
thing because $\nu_{\rm vertical}\approx \nu_{\rm orb}$ outside a neutron
star (for a discussion see Lamb \& Miller 2003).  There are details of
the observations (e.g., the conditions under which $\Delta\nu\approx
\nu_{\rm spin}/2$ instead of $\nu_{\rm spin}$) that are not obvious from
first principles (for some ideas see Lamb \& Miller 2003), but the general
constraints on models suffice to constrain masses and radii as long as
$\nu_{\rm upper}\approx \nu_{\rm orb}$ at some radius.

Titarchuk and colleagues (e.g., Titarchuk 2003) have suggested instead
that it is the {\it lower} peak frequency $\nu_{\rm lower}$ that is close
to $\nu_{\rm orb}$, with consequently different implications.  In their
model, the upper peak frequency $\nu_{\rm upper}$ is instead close to the
hybrid frequency $(\nu_{\rm lower}^2+4\nu_{\rm mg}^2)^{1/2}$, where the
magnetospheric frequency $\nu_{\rm mg}\approx\nu_{\rm spin}$.  This is an
interesting suggestion, but the recent high-precision measurements of
SAX~J1808 present puzzles for this model.  Chakrabarty et al. (2003) show
that the spin frequency is 401~Hz, rather than half this value. Wijnands
et al. (2003) find a pair of QPOs, at $\nu_{\rm lower}=499$~Hz and
$\nu_{\rm upper}=694$~Hz; the hybrid model would predict $\nu_{\rm
upper}=(499^2+ 4[401]^2)^{1/2}=945$~Hz, in conflict with the
observations. Similarly, data for KS~1731--260 present difficulties.
Wijnands \& van der Klis (1997) find $\nu_{\rm lower}=898$~Hz and
$\nu_{\rm upper}=1159$~Hz.  The burst oscillation frequency is 524~Hz
(Smith et al. 1997).  If $\nu_{\rm spin}=524$~Hz, the hybrid model
predicts $\nu_{\rm upper}=1380$~Hz.  If instead $\nu_{\rm spin}=262$~Hz,
the hybrid model predicts $\nu_{\rm upper}=1040$~Hz.  Both appear not in
accord with the data.  For this reason, we will concentrate on models in
which $\nu_{\rm upper}\approx\nu_{\rm orb}$.

In such models, measurement of $\nu_{\rm upper}$ constrains the mass and radius
of a neutron star.  If $R_{\rm orb}$ is the radius of a circular orbit
of frequency $\nu_{\rm orb}$, then clearly $R<R_{\rm orb}$.
In addition, the high quality factors of the QPOs require that they be 
produced outside the region of unstable
circular orbits predicted by general relativity.  For a nonrotating star,
for which the exterior spacetime is described by the Schwarzschild
geometry, the radius of the innermost stable circular orbit (ISCO) is
$R_{\rm ISCO}=6GM/c^2$.  The constraint $R_{\rm orb}>R$ places  a
mass-dependent limit on the radius; for example, for a nonrotating star
$R<(GM/4\pi^2\nu^2_{\rm orb})^{1/3}$ (Miller et al. 1998).
The additional constraint $R_{\rm orb}>R_{\rm ISCO}$ places an absolute
upper limit on the mass and hence on the radius.  When one considers
frame-dragging effects, the upper limits on the mass and radius are
(Miller et al. 1998)
\begin{equation}
\begin{array}{rl}
M&<2.2~M_\odot(1000~{\rm Hz}/\nu_{\rm orb})(1+0.75j)\\
R&<19.5~{\rm km}(1000~{\rm Hz}/\nu_{\rm orb})(1+0.2j)\; .\\
\end{array}
\end{equation}
Here $j\equiv cJ/GM^2$ is a dimensionless spin parameter, where $J$
is the stellar angular momentum.  If in a  particular case one
believes that the observed frequency is in fact the orbital
frequency at the ISCO, then the mass is equal to the upper limit
given in equation~(1).

The highest frequency QPO so far detected with
confidence has a frequency $\nu_{\rm QPO}=1330$~Hz (van Straaten et al.
2000), which would imply $M\lta 1.8\,M_\odot$ and $R\lta 15$~km for a 
system with spin parameter $j=0.1$.  These constraints essentially rule out 
the hardest equations of state proposed (see Figure~2).  The existence of the
ISCO means that the frequencies cannot be arbitrarily high.  If the
radius at which the QPOs are generated gets close to the ISCO, a variety
of signatures are possible, including flattening in the observed relation
between frequency and countrate, or sharp drops in the amplitude or
coherence of the QPO (see Miller et al. 1998).  
Such signatures would confirm the presence of unstable orbits, a key
prediction of strong-gravity general relativity, and allow a direct mass
measurement.  It is possible that the system 4U~1820--30 has already shown
such a signal (Zhang et al. 1998), but there are complications in the
spectral behavior that make this uncertain (M\'endez et al. 1999).

\begin{figure}
  \includegraphics[height=.3\textheight]{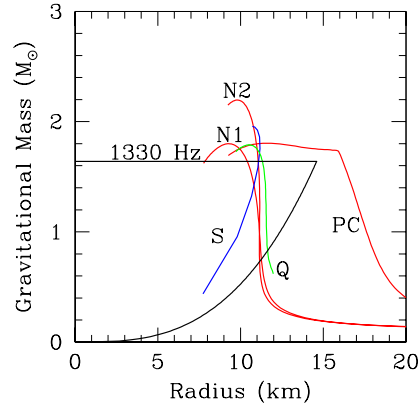}
  \caption{Constraints from orbital frequencies.  The 1330~Hz curve is for
the highest kilohertz quasi-periodic oscillation frequency yet measured (for
4U~0614+091, by van Straaten et al. 2000).  This curve is for a
nonrotating star; the constraint wedge would be enlarged slightly for a
rotating star (see Miller et al. 1998).  The solid lines are
mass-radius curves for different representative high-density equations of
state.  The mass-radius curves are all for equilibrium
nonrotating stars; note that rotation only affects these curves to second
order and higher. Curves N1 and N2 are for nucleonic equations of state; N1
is relatively soft (Friedman \& Pandharipande 1981),  whereas N2 includes
significant three-body repulsion (Wiringa, Fiks, \& Fabrocini 1988).   PC
has a sharp change to a Bose-Einstein condensate of pions in the core when
the mass reaches $\approx 1.8\,M_\odot$ (Pandharipande \& Smith 1975).
Equations of state N1, N2, and PC are not modern (i.e., not fitted to the
most current nuclear scattering data), but are included for easy comparison
to  previous work on equation of state constraints. Curve S is for a
strange star equation of state (Zdunik 2000).  Curve Q is a quark
matter equation of state with a Gaussian form factor and a diquark
condensate (kindly provided by David Blaschke and Hovik Gregorian).}
\end{figure}

\section{Prospects With a 10~m$^2$ Instrument}

As discussed by Swank (these proceedings), a trend evident from RXTE
observations is that as the mass accretion rate increases (inferred, e.g.,
from the $S_a$ index; M\'endez et al. 1999),  the frequencies of QPOs
increase and the amplitudes of QPOs decrease.  For many sources observed
with RXTE, the inferred mass accretion rate continues to increase after the 
QPOs become unobservable.  This suggests that a larger area timing instrument
would be able to detect higher frequencies.  For example, for 4U~0614+091,
projection of the amplitude versus accretion rate trends suggests that a
10~m$^2$ instrument could detect frequencies of $\sim 1500$~Hz (M. van der
Klis, personal communication).  Other atoll sources could yield even higher
frequencies.

An important threshold is reached at $\sim 1500$~Hz, because this is at or
above the orbital frequency at the ISCO for realistic masses.  The
orbital frequency at radius $r$ in a Kerr spacetime of spin parameter
$j$ is $\Omega=M^{1/2}/(r^{3/2}+jM^{3/2})$ (e.g., Shapiro \& Teukolsky
1983, equation 12.7.19), and to ${\cal O}(j)$,
$R_{\rm ISCO}=6M[1-(2/3)^{3/2}j]$ (see, e.g., Miller \& Lamb 1996),
so the orbital frequency is
\begin{equation}
\nu_{\rm orb,ISCO}\approx 2199~{\rm Hz}(M_\odot/M)
\left[1+{11\over 8}\left(2\over 3\right)^{3/2}j\right]\; .
\end{equation}
For $j=0.1$, $\nu_{\rm orb,ISCO}<1500$~Hz for $M>1.58~M_\odot$. It has
long been suspected that neutron stars in LMXBs are more massive than the
canonical $1.4\,M_\odot$ because of mass transfer, and direct evidence of
this has arrived recently.  Nice, Splaver, \& Stairs (2003) report
that the 22~ms pulsar in the 0.26~day binary J0751+1807 (with a low-mass
white dwarf companion, likely a remnant after substantial mass transfer) has
$M>1.6\,M_\odot$ at better than 95\% confidence, and the mass could be
well above this.  Measurements of QPOs above 1500~Hz therefore have
excellent prospects for stronger constraints on masses and
radii, and even for detection of signatures of the ISCO.  A QPO frequency as
high as 1800~Hz would be large enough to argue against all standard
nucleonic or hybrid quark matter equations of state, leaving only 
strange stars (see Figure~3).

\begin{figure}
  \includegraphics[height=.3\textheight]{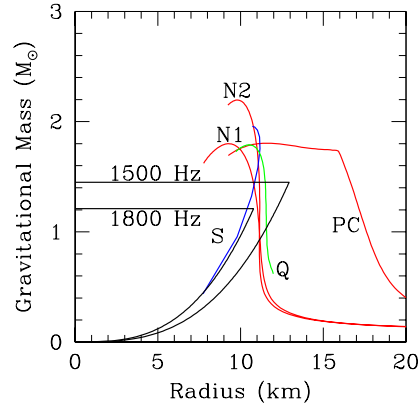}
  \caption{Constraints on mass and radius for hypothetical detections
of a 1500~Hz QPO and a 1800~Hz QPO, if they are identified with an
orbital frequency.  At 1500~Hz one expects signatures of the ISCO to
be present; a detection of 1800~Hz would present strong difficulties
for standard nucleonic equations of state.  The equation of state curves
are as in Figure~2.  Rotational effects are not included in this figure.}
\end{figure}

A qualitative advantage of a large-area timing instrument compared with RXTE is
that there are a number of sources with strong enough QPOs that they
could be detected in less than a coherence time.  For example, in some
states, the kHz QPOs in Sco X-1 could be detected within $\approx$4~ms
with a 10~m$^2$ instrument (M. van der Klis, personal communication).
Because current detections are averaged over many coherence times, it is
not possible to determine whether, e.g., QPOs are present at all times
or whether they are superpositions of more coherent pulses.  A larger
area instrument would help resolve this, and if indeed there are
underlying highly coherent pulses this could lead to substantial
additional insights.  For example, if the QPOs are caused by the orbits
of radiating clumps, then observation within a coherence time would lead
to detection of Doppler shifts, which when combined with the observed
frequency would allow a unique solution of both the gravitational mass
of the neutron star and the radius of the orbit.

In conclusion, RXTE observations have not only revealed a previously
unsuspected phenomenon, but have constrained models of kHz QPOs 
significantly.  With current data, we are just short of expected
signatures of the innermost stable circular orbit, a crucial predicted
characteristic of strong gravity.  A larger-area instrument is likely
to push us over this important threshold, and also to allow novel new
methods of analysis that can detect qualitatively new phenomena such
as periodic Doppler shifts of orbiting clumps.

\begin{theacknowledgments}
  We are grateful for many useful discussions with Fred Lamb, Dimitrios
Psaltis, Jean Swank, and Michiel van der Klis, and for the organizational
efforts of Phil Kaaret, Fred Lamb, and Jean Swank.  This work was supported 
in part by NSF grant AST~0098436.
\end{theacknowledgments}

\bibliographystyle{aipproc}   % if natbib is available

\end{document}